\documentclass[journal, a4paper,twocolumn]{IEEEtran}
\IEEEoverridecommandlockouts 
\ifCLASSINFOpdf
\usepackage[pdftex]{graphicx}
\graphicspath{{./Images/}}
\DeclareGraphicsExtensions{.pdf,.jpeg,.png}
\else
\fi

\usepackage[cmex10]{amsmath}
\usepackage{siunitx}
\usepackage{cite}
\usepackage{psfrag}
\usepackage[utf8]{inputenc}
\usepackage[T1]{fontenc}
\usepackage{amsmath,amsfonts,amsbsy,amssymb}
\usepackage{mathabx}
\usepackage{mathrsfs}
\usepackage[nolist]{acronym}
\usepackage{tabularx}
\usepackage{amssymb}
\usepackage{amsmath}
\usepackage{graphicx}
\usepackage{cite}
\usepackage{multirow}
\usepackage{wasysym}
\usepackage{multirow}
\usepackage{float}
\usepackage{color}

\newcommand{\new}[1]{\textcolor{black}{#1}}
\newcommand{\newii}[1]{\textcolor{black}{#1}}

\begin{document}
\title{Demand Control Management in Micro-grids: The Impact of Different Policies and Communication Network Topologies}

\author{Florian Kühnlenz, Pedro H. J. Nardelli, Hirley Alves
\thanks{The authors are with the Centre for Wireless Communications (CWC) at University of Oulu, Finland. P. H. J. Nardelli is also with Laboratory of Control Engineering and Digital Systems, Lappeerntanta University of Technology, Finland. This work is partly funded by Finnish Academy (n. 271150) and CNPq/Brazil (n.490235/2012-3) as part of the joint project SUSTAIN, by SRC/Aka BCDC Energy (n.292854) and by the European Commission through the P2P-SmarTest project (n.646469). Contact: pedro.nardelli@lut.fi}%
}

\maketitle

\begin{abstract}
This work studies how the communication network between proactive consumers affects the power utilization and fairness in a simplified direct-current micro-grid model, composed by three coupled layers: physical (an electric circuit that represents a micro-grid), communication (a peer-to-peer network within the micro-grid) and regulatory (individual decision strategies).
Our results show that, for optimal power utilization and fairness, a global knowledge about the system is needed, demonstrating the importance of a micro-grid aggregator to inform about the power consumption for different time periods.
\end{abstract}

\begin{IEEEkeywords}
Communication networks, Agents-based systems, Network topology, Smart grids, Decision making.
\end{IEEEkeywords}


\section{Introduction}
The growth of communication networks within the electric power grid allows for different new applications to its end-users \cite{Bera2014}. 
This, on its turn, increases the complexity involved in the management of such a system.
In this context, there is a need to develop models that aim at capturing new features that might emerge \cite{nardelli2014models}, {as for example synchronization of individual behaviors due to signals and incentive structures.}
{Game theory illustrates one approach that can capture such features since it is able to model strategic interactions of individual agents.
It is worth mentioning that this theory was firstly introduced to model distributed economic relations, but is now widespread within other research communities (e.g. \cite{Mohsenian-Rad2010,Wang2014} and references therein).}

The present work focuses on electricity micro-grids (specific region that can be self-sustained in relation to the large power grid, e.g. isolated rural areas with local energy generation)  and is related to the numerous studies done on the fields of economics \cite{Saad:2012jx}, power systems \cite{Engels:2016ep,Almasalma:2016ti} and communication systems \cite{safdar2013survey}.
Here we follow the ideas presented in \cite{bale2015energy}, where the authors pointed out some limitations of the standard economic models for energy systems while indicating the importance of complexity science approaches.
Like some articles as \cite{Engels:2016ep,Almasalma:2016ti,Alvarado:2001,SimpsonPorco:2013gs}, we incorporate here aspects from different fields {to include how peer interactions or collective individual actions may affect the macro-level outcomes as in gossip-based algorithms or electricity markets.}

To do so, we use an agent-based model, recently proposed by the authors in \cite{Nardelli2015a}, to model agents' interactions within a system constituted of three layers (physical, communication and regulatory, to be explained later) and then obtain a better understanding of both the inter- and intra-layer dynamics.
We therefore expect to provide a different perspective about demand-side management policies \cite{gottwalt2011demand,Callaway2011,ZhangSJ2016,PradhanSJ2016} that goes beyond individual utility maximization and purely competitive strategies (usual assumptions of many economic researches \cite{bale2015energy}).

\newii{In specific terms, we extend our previous model  of a fully distributed system operation by introducing an entity that provides a centralized signal.
This was motivated by the demand control problem in direct current (DC) micro-grids.
We consider three different scenarios related to how information is shared among consumers: no signaling (similar to \cite{Nardelli2015a}), global state signaling and pricing scheme.
The first strategy is fully distributed, while the other two require an entity with global state information about the whole system; this entity is the micro-grid aggregator.
Our main contribution is to assess these different demand-side management policies  (without or with micro-grid aggregator) in both micro and macro levels looking at the agent decision dynamics and considering different communication network topologies.
It is worth mentioning the present paper is the first attempt to study micro-grids  using the model introduced in \cite{Nardelli2015a}.}

Our model is described as follows.
The physical layer (a DC micro-grid) is a circuit composed by a power source and resistors in parallel.
Individual agents (the proactive consumers, the ``prosumers'') can add, remove or keep the resistors they have. 
Agents' decisions aim at maximizing their own delivered power, which is a non-linear function dependent on the others' behavior, and they are based on (i) their internal state, (ii) their global state perception,  (iii) the information received from their neighbors in the communication network, and (iv) a randomized selfishness related to their willingness of answering to a demand-side control request.
Different peer-to-peer communication network topologies and randomized communication errors in the deployment of micro-grids are important aspects to be included as in \cite{li2014joint,khan2016cognitive}. 
But differently from other papers, we analyze here their systemic effects; for detailed results of   communication network implementation  in power grids using, for instance, spatial and/or temporal spectrum sharing, refer to \cite{nardelli2016maximizing,tome2016joint,cacciapuoti2016probabilistic,cacciapuoti2015enabling,Cacciapuoti2016}.

Looking at the proposed model, by individually modifying the peer-to-peer communication network topology and the demand-side management policies keeping fixed the other parameters including the link error probability, we expect to show how these factors affect the power utilization and fairness in the system for different number of agents.
{Note that our study assumes that each one of those three layers  (physical, communication and regulatory) is equally important to determine the system dynamics and the individuals' (re)actions.}
In this case, the system can neither be reduced to one or two of these layers nor the dynamics of the whole system cannot be assessed based on individual (typical) agents.

Our previous results in \cite{Nardelli2015a} can be seen as the benchmark scenario where no demand-side management is considered; there we showed: (i) different communication network topologies (ring and Watt-Strogatz-Graph \cite{boccaletti2006complex}) lead to different levels of power utilization and fairness at the physical layer and (ii) a certain level of error induces more cooperative behavior at the regulatory layer.
Now, if demand-side control or pricing schemes are considered, which is the focus of the present work, the system behaves in a more predictable manner and is less dependent on its size.
In this case, the global knowledge about the system state enables much higher utilization and fairness, which indicates the need of an entity to guide the peer relations.
It is important to say that, although there is a wide understanding that global signaling leads to more stable outcomes, this is not always the case; many times global signaling induces destabilizing collective behavior (e.g. \cite{nardelli2017smart} and references therein).

The rest of this paper is divided as follows.
Section \ref{sec:model} describes the multi-layer model employed here.
Section \ref{sec:results} presents the numerical results used to analyze the proposed scenario.
In Section \ref{sec:final}, we discuss the lessons learned from our model, indicating potential future works.
The most important symbols are presented in Table \ref{table:notations}.

\begin{table}[!t]
	\caption{Notations}
	\label{table:notations} \centering
	\begin{tabular}{l|l}
		\textbf{Notation} & \textbf{Meaning}   \\\hline
		$t \in \mathbb{Z}$ & discrete time \\ 
		$\mathcal{A} = \{1,2,...,N\}$ &  set of agents; $N$ is the system size\\ 
		$\mathcal{A}\backslash \{i\}$ & set $\mathcal{A}$ without agent $i$\\
		$i \in \mathcal{A}$ & agent $i$\\ 
		$\mathcal{N}_i \subset \mathcal{A}$ &  neighborhood set of agent $i$\\ 
		$n[t] \in \{N,N+1,...\}$ &  active number of resistors at  $t$\\ 
		$a_i[t] \in \mathbb{N}^+$ &  active resistors of agent $i$\\ 
		$r_i[t] = n[t] - a_i[t]$ &  active resistors excluding agent $i$\\ 
		$P_i[t] > 0$ &  agent $i$ consumed power [units of power]\\ 
		$\lambda_i[t] \in \mathbb{R}$ &  gain in power of agent $i$\\ 
		\new{$\lambda_\mathrm{min} \in \mathbb{R}$} & \new{system-wide predefined minimum gain}\\ 
		$S_i[t] \in \{-1,0,+1\}$ &  agent $i$'s state: $+1$ (defect),\\ & $0$ (ignore), or $-1$ (cooperate)\\ 
		$s_i \in [0,1]$ &  selfishness gene of agent $i$\\ 
		$p_{\mathrm{err}} \in [0,1]$ & error probability in a communication link\\
		%
		$c_avg \in [0,1]$ & average value of cooperating agents\\
		\new{$R$} & \new{resistor value added as a load by agents}\\
		\new{$R_\mathrm{V}$} & \new{source resistor}\\

	\end{tabular}
\end{table}

\section{Multi-layer model}
\label{sec:model}

\subsection{Background}

The power grid has been thought as a one-way network, where the generators produce the electricity that needs to travel long distances in high-voltage transmission lines to arrive at the distribution network, which delivers the electricity to final consumers \cite{nardelli2014models,Nardelli2015a}.
However, in order to integrate ever increasing wind and solar power, this network has to change in many ways.
First, solar and wind energy need more physical space than classical power plants, since they are not just transforming energy from one high density form to another, but rather collecting low density energy while distilling it in to a high density form.
Second, they will produce energy in a much less predictable and/or consistent way.

\new{This volatility in supply means that the consumption and distribution need to react to these changes, either though storing energy or shifting consumption in time.}
In this case, automation is needed to balance the system; therefore, the so-called smart grids are planned to heavily use information and communication technologies (ICTs) to collect and take autonomous actions. 
Communication technologies should then enable the grid to use electricity from diverse sources whenever they are available, delivering to wherever they are needed.
This would improve the efficiency of both production and distribution since the smart grid is designed to provide the necessary capabilities to enhance the reliability of energy infrastructures, and decrease the impact of supply disruptions, as well as create new services and applications. 
To achieve the needed degree of automation to integrate volatile energy sources, the power grid needs to be coupled with the communication network. 
This creates a codependency, where the power grid will only work when the communication system is working, which in turn will only work if the power system is working.
Furthermore, the volatility of the power sources demands a new interaction paradigm between producers and consumers.
Consumers need to react to changes in the supply, by shifting their demand, whenever possible, to synchronize with production.
This means that the smart grid will also be much more coupled to a regulatory layer, which will influence consumer behavior and decisions.

In summary, more than any specific technological challenge existing in the generation, transmission and distribution, a proper smart grid model should include the following different layers of analysis. 
\begin{itemize}
	%
	\item \textbf{Physical electricity grids} that include intermittent sources of energy (e.g. solar panels or wind turbines), micro-grids and mobile batteries/loads (e.g electric vehicles). The energy interchange occurs in this layer, transported in form of electricity.
	%
	\item \textbf{Information networks} to acquire, process and disseminate information respecting the specific application requirements. The communication between agents happens in this layer, which can be seen as the agent reality (i.e. the agent only knows about itself and whatever is communicated to it).
	%
	\item \textbf{Regulatory layer or markets} where consumers and operators can interchange electricity, evolving different strategies under different exchange regulations. This layer therefore include the socio-economical context and decision-making processes, which directly affects the physical layer.
	%
\end{itemize}

\begin{figure*}
\centering
\includegraphics[width=2\columnwidth]{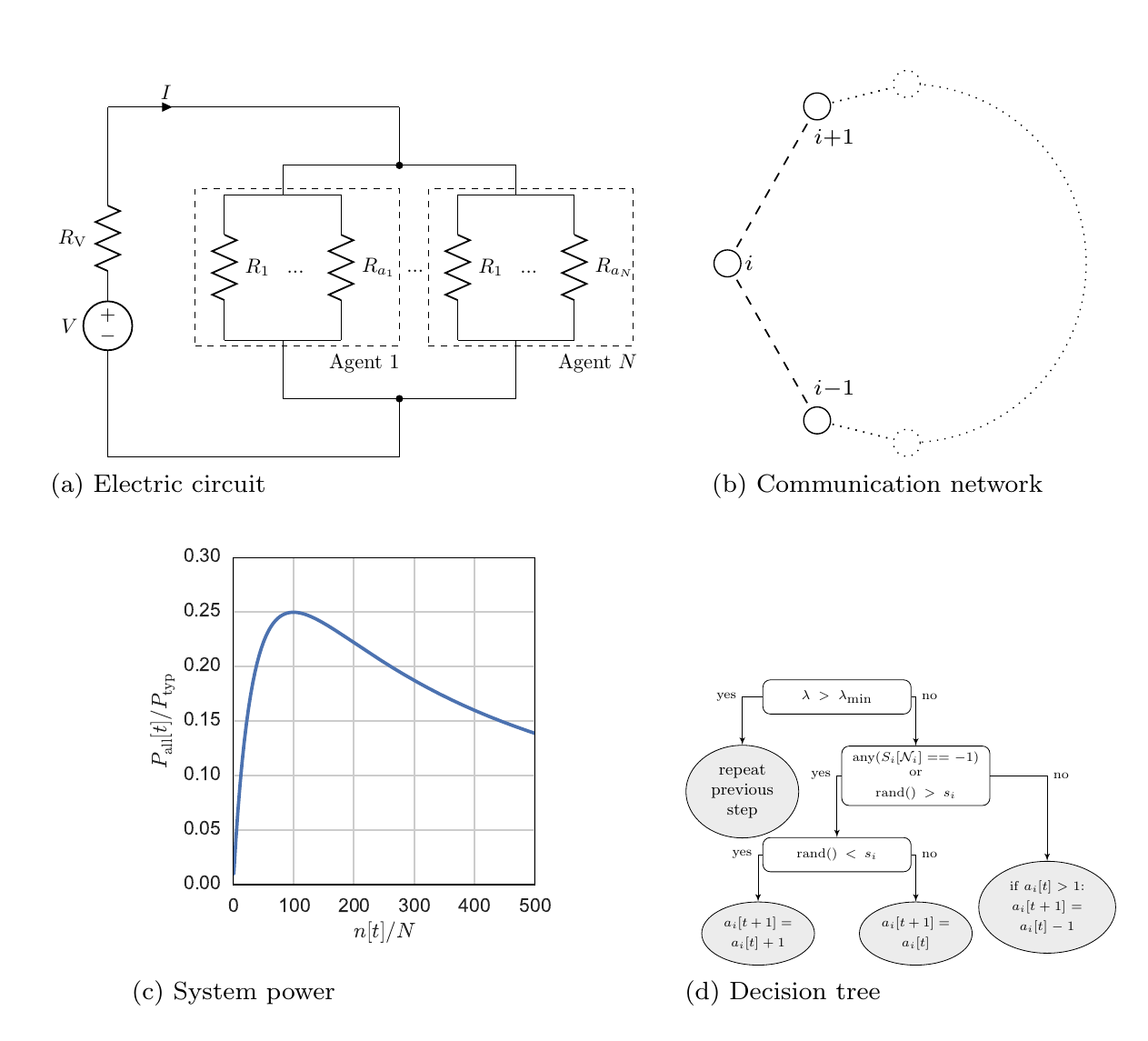}
\caption{(a) Electrical circuit representing the physical layer of the system. The circuit is composed by a power source $V$ and $R_\mathrm{V}$, and resistors of $R$ in parallel, generating a current $I$. These resistors are related to $N$ agents that can add, remove or keep the resistors under their control in the circuit.
The minimum number of resistors an agent can have is one and there is no maximum.
We also consider $N$ as the size of the system.
(b) The agents are connected in a communication network so that a given agent has access to the information related to the previous action of their first-order neighbours. In the ring topology illustrated here, every agent is connected with two other agents. In this case agent $i$ is connected with agents $i-1$ and $i+1$ with $i = 1,2,...,N$. In the ring topology, agents $1$ and $N$ are neighbors.  
(c) At time step $t$, the normalized power delivered  $P_\mathrm{all}[t]/P_\mathrm{typ}$ to the agents with rising number of resistors $n[t]$ in the system, where $P_{\mathrm{typ}}=\frac{V^{2}}{R_\mathrm{V}}$ and $P_\mathrm{all}[t] = \sum_{i\in \mathcal{A}} P_i[t]$ with $P_i[t]$ given by equation \eqref{eq-power-i}.
(d) The decision tree representing the decision making process of each agent for each round.}
\label{fig:System} 
\vspace{2ex}
\end{figure*}

Instead of analyzing each one of them separately, considering the others as given (the most usual approach), this research will  model these three layers as constitutive parts of the smart grid system.
By doing so, we will be able to provide a deeper understanding of the smart grid spatio-temporal multi-layer dynamics and the emergent systemic features that arise from them. 
It is important to say that this proposal is built upon the assumption that effective cross-layer management strategies targeting the realization of the ambitious goals planned for the smart grids can be only realized if the way that the system elements interact within one layer and across different layers is properly characterized.

This approach is philosophically inspired by the Russian/Soviet tradition of reflexive systems and control \cite{Lepskiy2017}.
Such a perspective in their current version has two fundamental aspects captured by our model:
the importance of the information layer as the filter of the physical reality, and the reflexive effects of agents and contexts that they take their decisions.
In other words, what individual agents perceive is as important as (or even more important than) what is actually happening in the physical layer.
Their individual decisions at the regulatory layer, in turn, depend on a combination from their own internal state and their own perceptions of the system state.
The individual decisions, which are built upon different perspectives,  will determine the physical layer dynamics, which shall affect the future decisions of individual agents (i.e. reflexive behavior).
In practical terms, the present work is a multi-layer model of DC micro-grids, as part of smart grids, that considers also peer-to-peer exchange of information within a network with different topology; this model will be briefly described in the following.
Our discrete-time, agent-based model assumes these three layers as constitutive parts of the system composed by an electric circuit as the micro-grid physical infrastructure, a communication network where agents exchange local information and a set of regulations that define the agents' behavior, as exemplified in Fig. \ref{fig:System}.
It is worth saying that this model -- as any other model -- is a clear simplification of the real grid.
Our goal here is to put light on the systemic multi-layer dynamics, which is most often neglected.

\subsection{Agents' decision process}
As stated before the model assumes discrete time steps, denoted by $t$.
Therefore, the interactions between agents might be viewed as a round-based game \cite{archetti2012review} based in the state $S_i[t]$, defined at Table \ref{table:notations} and discussed later on.
At each step in time $t$, every agent aims to maximize its own consumed power.
To achieve this goal, the agent has three options:
\begin{itemize}
	\item \textbf{Add a resistor $R$} (defecting); 
	\item \textbf{Remove a resistor $R$} (cooperating); or 
	\item \textbf{Do nothing} (ignoring).
\end{itemize}
\new{We assume that the value of $R$ is fixed and the same for all agents.
Our preliminary investigation indicated that relaxing this assumption within a reasonable range of values has no significant effects on the system dynamics; then, we decided to use the homogeneous case for simplicity.}

The decision is based on the gain from the previous strategy $S_i[t-1]$ in order to decide its new state $S_i[t]$ in the following manner.
If the gain $\lambda_i[t-1]$ is greater than or equal \new{to a system-wide predefined minimum $\lambda_\mathrm{min}$}, the agent continues to its (successful) strategy at time $t$, i.e. $S_i[t] = S_i[t-1]$.

If $\lambda_i[t]<\lambda_{\mathrm{min}}$, then agent $i$ compares its  strategy with its neighborhood $\mathcal{N}_i$, which is related to the communication network and will be defined later in this section.

In the case that a majority of the neighborhood is cooperating, e.g. $\sum_{j \in \mathcal{N}_i} S_j[t-1] < 0$, this will appear to agent $i$ as an indication that the system is in a condition of overusing.
The agent will then also change to a cooperative state, leading to $S_i[t] = -1$.
Otherwise, the agent decision will be related to its individual selfishness as follows: a random number between 0 and 1 is drawn to be compared to its own selfishness gene $s_i$ in order to decide whether it will start cooperating or not.
Each agent is assigned with selfishness gene at the beginning of the simulation, which is also a random number between 0 and 1.
In the case of not cooperating, another random number is drawn and once again compared to the selfishness gene $s_i$, but now to decide if stays inactive (i.e. $S_i[t] = 0$) or adds another load in the circuit (i.e. $S_i[t] = +1$). 
The agent decision procedure is shown in Fig. \ref{fig:System} (d).

\subsection{Communication network}
The communication network enables agent $i$ to know the state $S_j[t-1]$ of the agents $j \in \mathcal{N}_i$ which are in his neighborhood.
We assume that agent $j$ always transmits its actual state $S_{j}[t]$ to agent $i$.
However, the communication links can also experience errors.
An error event means the received message by agent $i$ contains a different information than agent $j$ has sent.
So that if $S_{j\rightarrow i}[t-1] = S_{j}[t-1]$ is the state information send from $j$ to $i$ and $\hat{S}_{j\rightarrow i}[t-1]$ be the information received by $i$.

We assume that error events are independent and identically distributed (i.i.d.) such that the probability of the event: $\mathrm{Pr}\left[\hat{S}_{j\rightarrow i}[t-1] \neq S_{j\rightarrow i}[t-1] \right] = p_\mathrm{err}$ for all $t \in \mathbb{Z}$, $i \in \mathcal{A}$ and $j \in \mathcal{N}_i$, $\mathrm{Pr}[\cdot]$ refers to the probability that a given event occurs.
The network is a bidirectional graph so that an error at $i \rightarrow j$ does not imply an error at $j \rightarrow i$, and vice-versa.
If an error happens, the received information $\hat{S}_{j\rightarrow i}[t-1]$ is assumed to be i.i.d. between the other possible states.

In this study we compare two different communication networks \cite{boccaletti2006complex}:  
\begin{itemize}
	\item \textbf{Ring:} Each agent connects to exactly two other nodes to form a ring. It is a simple deterministic topology. The agents have then the same degree.
	\item \textbf{Watt-Strogatz-Graph:} Random graph that has small-world properties, namely short average path lengths and high clustering. In this case, the agents have, in average, similar degrees. 
\end{itemize}
\new{We did not consider here Barabasi-Albert graphs since the scenario under investigation presents a similar macro-level behavior to Watts-Strogatz, as shown in \cite{Nardelli2015a}}.

\subsection{Physical system}
In the physical systems as presented in Fig. \ref{fig:System}, there exists an optimal number of resistors that leads to the maximum power delivered from the source to the agents.
If the delivered power is below the maximum on the right, then there will be a gain by removing a resistor until the system has reached such point.
Conversely, if it is below on the left, then there will be a gain by adding a resistor.

\begin{figure*}[!ht]
	\centering
	\includegraphics[width=2\columnwidth]{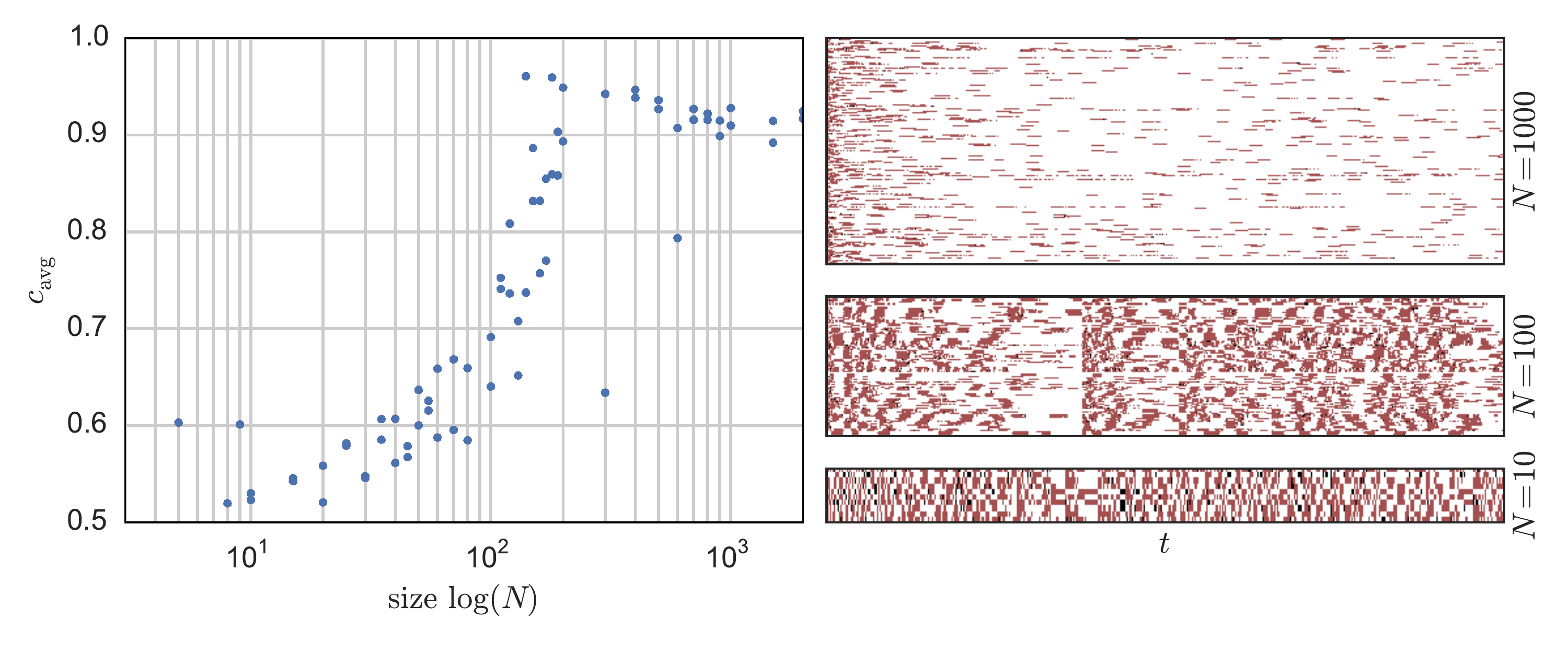}
	\caption{On the left: change
		in the average cooperation $c_\mathrm{avg}$ depending on the system size $N$. On the right: examples of a typical system
		behavior as a function of time $t$ where the points (pixels) represent the agent state $S_i[t]$ such that red means defection, white cooperation, and black =
		doing nothing for $N=1000$ (top), $N=100$ (middle) and $N=10$ (bottom). The system parameters are: $\lambda_\mathrm{min}=0.0005$, $R_V=\SI{2}{\ohm}$, $R_0= R/N =\SI{200}{\ohm}$, $p_\mathrm{err}=0.01$ and $V=\SI{1}{\volt}$. The communication network is configured as Watts-Strogatz graph.
		\label{fig:transition-WS}}
\end{figure*}

One may ask the following question: \textit{Is it possible for the agents to reach the optimal point with limited knowledge about the system?}
In the presence of a central controller this would be a fairly easy problem.
First, we need to find the number of resistors that leads to optimal power to then distribute them among the agents.
This kind of solution resembles time division schemes in computer networks or cellular systems \cite{marsic2010computer}.
However, as discussed before, in the absence of a centralized controlling entity, the agents only have a limited knowledge about other agents.

In any case, the agents have some information about the state of the system based on their own power consumption and the decision done.
At time-step $t$, the power each agent consumes $P_i[t]$ with $i\in\mathcal{A} = \{1,...,N\}$ and is given by:
\begin{equation}
\label{eq-power-i}
P_{i}[t]=P_{\mathrm{typ}}\frac{a_i[t] \, \mu}{(a_{\mathrm{avg}}[t]+\mu)^{2}},
\end{equation}
where $P_{\mathrm{typ}}=\tfrac{V^{2}}{R_\mathrm{V}}$, $\mu=\tfrac{R}{R_{\mathrm{V}}}$, $a_i[t]$ is the number of active resistors the agent $i$ possesses, $r_i[t]$ is the number of active resistors in the system excluding the source resistor $R_\mathrm{V}$ and the ones controlled by agent $i$, and $a_{\mathrm{avg}}[t]=(a_i[t]+r_i[t])/N$.
The physical system is then described by its size $N$, the ratio $\mu$ of the resistance values and the power source $V$.
The resistors are scaled so that the optimal average number of resistors $\left(a_{\mathrm{avg}}^{*}\right)$ is independent of $N$ while the voltage might be scaled with $\sqrt{N}$ to have a constant ratio of power per agent, as explained later on.
The gain that agent $i$ experiences at time-step $t$ is then defined as:
\begin{align}
\label{eq-gain-i}
\lambda_i[t]&=\frac{P_{i}[t]-P_{i}[t-1]}{P_{i}[t-1]} \nonumber \\
&= \dfrac{\Delta P_i}{P_{i}[t-1]}.
\end{align}

This implies that the agents only use the information about the previous time-step $t-1$.
If we expand \eqref{eq-gain-i} using  \eqref{eq-power-i}, the resulting equation that determines $\lambda_i[t]$ becomes more complicated.
To make the analysis clearer, we choose to apply the following approximation:

\begin{align}
\lambda_i[t] &\approx \frac{\mathrm{d} P_{i}}{P_{i}[t]} \nonumber \\
&\approx\frac{\Delta a_i[t]}{a_i[t]}-\frac{\;2\;}{N} \; \frac{1}{a_{\mathrm{avg}}[t]+\mu}\left(\Delta r_i[t]+\Delta a_i[t]\right),
\label{eq:gain}
\end{align}
such that the gain $\lambda_i[t]$ is now a function of the variations in agent $i$'s own number of resistors $\Delta a_i[t] = a_i[t]  - a_i[t-1]$ and in the number of resistors controlled by other agents $\Delta r_i[t] = r_i[t] - r_i[t-1]$,  as well as the average number of resistors $a_{\mathrm{avg}}[t]$ and the system parameters $N$ and $\mu$.

\section{Numerical results}
\label{sec:results}

In this section, we present our main results.
\new{We first revisit the results introduced in \cite{Nardelli2015a} where no demand-side management signaling is considered and the agents try to achieve a global optimum solely based in their local information.}
We then present our new results evaluating the effects of such factors on the system dynamics in comparison with our baseline model.
Once again, it is important to say that the numerical results to be presented only exemplify the impact of different ways of signaling on the multi-layer system dynamics. Although qualitative changes are expected to happen consistently, the actual numerical values depend on the actual signal construction.

\subsection{Baseline model}
\label{subsec:baseline}

In Fig. \ref{fig:transition-WS} one can see the inherent dynamics of the system when no demand-side management is done and the communication network topology is a random graph. 
We identify two phenomena that are important for our analysis, as explained next. 

\begin{figure*}[t]
	\centering
	\includegraphics[width=2\columnwidth]{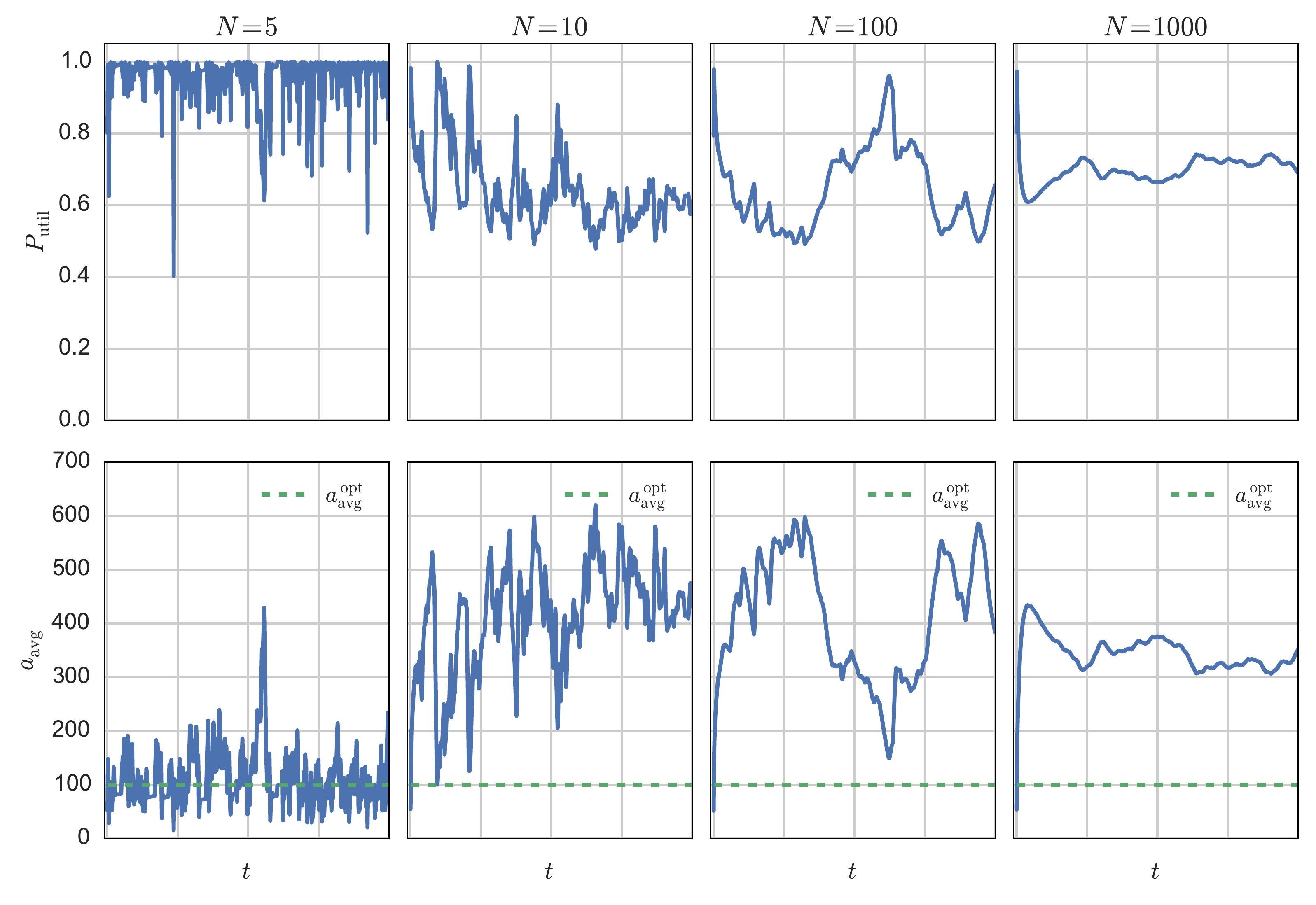}
	\caption{Typical behavior of very small ($N=5$), small ($N=10$), medium ($N=100$) and large systems ($N=1000$). The small systems can operate much closer to the optimum, while mid-size systems show big jumps after stable periods. The system parameters are  $\lambda_\mathrm{min}=0.005$, $R_{\mathrm{V}}=\SI{2}{\ohm}$, $R_0= R/N =\SI{200}{\ohm}$, $p_\mathrm{err}=0.01$ and $V=\SI{1}{\volt}$. The communication network is configured as a ring. \label{fig:behavior-plots}}
\end{figure*}

First, we see that the average number of cooperators varies with the size of the system, showing a very high number of cooperation for larger systems. 
\new{However, as we will see later in Fig. \ref{fig:transition-WS}, this is not sufficient to understand the system dynamics.}
This high level of cooperation is not necessarily a universally good outcome.

The second phenomena is a change in behavior of the system for different sizes.
While for \textit{small systems} we see an almost checkerboard like distribution of  white (cooperation) and black (defection) dots, \textit{mid-sized systems} show a strikingly different image, in this case we now see a wave-like pattern where a large front of cooperation establishes and then suddenly brakes down. 
This is again in contrast to the picture for \textit{large systems} where the behavior is dominated by cooperation with some sparsely distributed red dots.

The resulting changes in the power delivery can be seen in Fig. \ref{fig:behavior-plots}, which shows that only very small systems of less then 10 agents are able to stay very close to the optimum, while systems of more then 10 agents start to deviate from it. 
These curves also reflect the wave-like behavior for mid-sized systems, where the number resistors drops due to almost global cooperation.
This brings the system close to the optimal point, when the number of resistors rises once again. 
For large systems, it does not get close to the optimum.
However, it is more stable and predictable than the other cases.

The background of how this global behavior emerges from the local interactions is found in our previous work \cite{Nardelli2015a}. 
Here, we will view this intrinsic  unstable performance as a undesirable outcome.
In the next subsection, we will explore ways to prevent this behaviors by using demand-side signaling with different communication topologies.
As we will see later, our goal is to provide a global indication via either direct signaling or pricing so the agents can use this information to make their individual decisions.

\subsection{Demand-side management and the micro-grid aggregator}
\begin{figure}[!t]
	\centering
	\includegraphics[width=1\columnwidth]{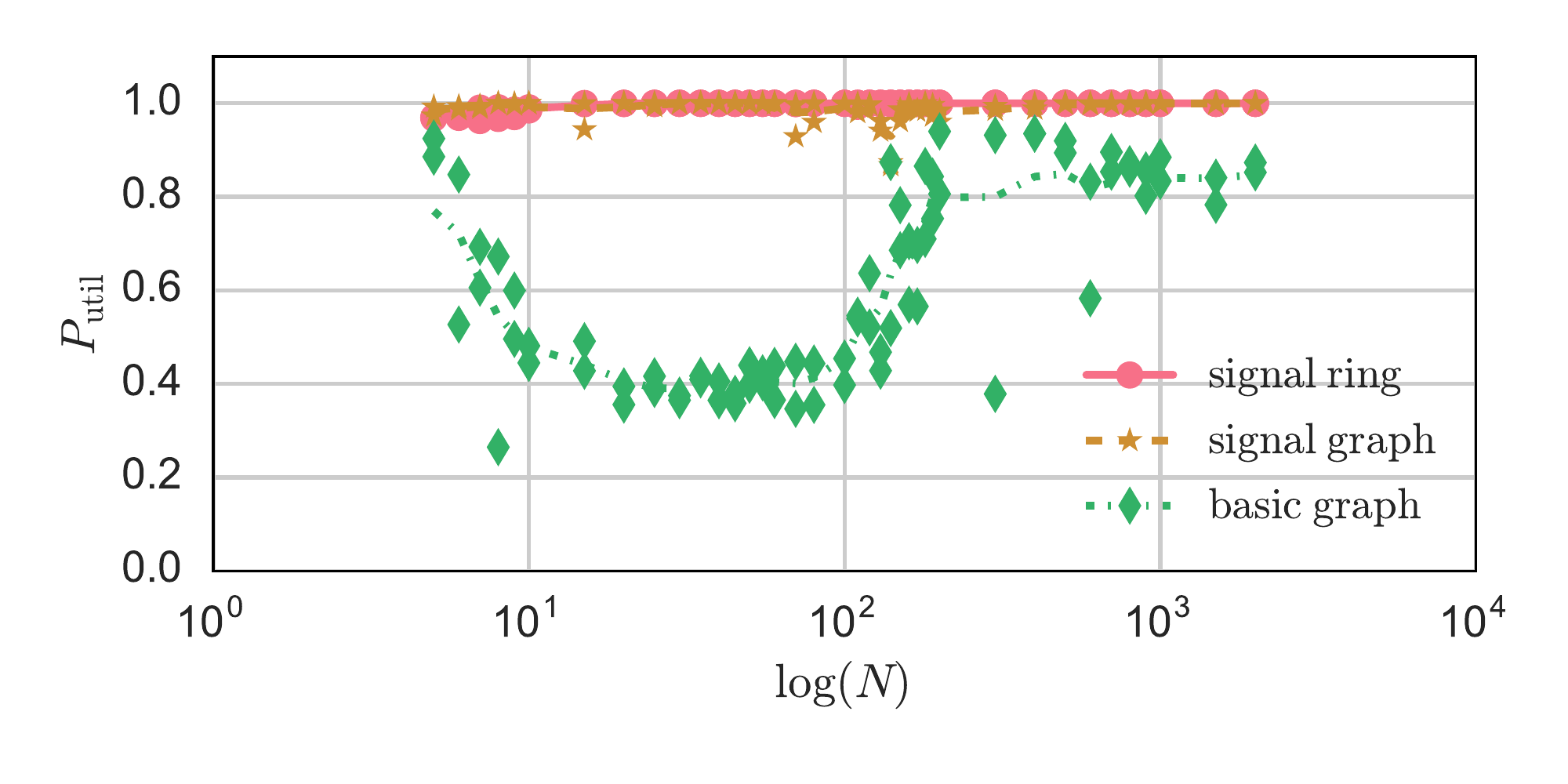}
	\caption{\label{fig:utilI}Power utilization for different communication strategies and topologies depending on the system. The green line shows the utilization for the basic model. The red and brown lines are from the simulations with global demand response signal.}
	%
\end{figure}

\new{Let us first define the power utilization $P_{\mathrm{util}}$ as the fraction of power that is utilized by the system and the available power:
\begin{equation}
	P_{\mathrm{util}}=\frac{4}{P_{\mathrm{typ}}} \;\sum_{i \in \mathcal{A}}P_{i,\mathrm{avg}},
\end{equation}
where $P_{i,\mathrm{avg}}$ is the time average power of agent $i$ computed as:
\begin{equation}
	P_{i,\mathrm{avg}}=\; \lim\limits_{T \rightarrow \infty} \; \dfrac{1}{T} \sum\limits_{t=0}^{T-1} P_i[t].
\end{equation}
}
\new{The first demand-side management policy is based on a global signal that every agent receives and labeled signal in the figures.} 
In Fig. \ref{fig:utilI}, we can see the results of such a global signal that is sent to all agents when the system is beyond the optimal point.
\new{The results are shown in comparison to the base case, labeled ``basic graph'' as presented in the previous subsection.}
\new{This signal might be interpreted as a way for a micro-grid operator to implement a demand response mechanism when the system is experiencing a peak in consumption, while supply is very limited}.
\new{Such a system are already found in the reality. As discussed in \cite{yadoo2011low}, the Scottish Island Eigg uses a traffic light system to signal its inhabitants the state of supply in the micro-grid.}

From the agents' view, this signal provides a global information about the system so they do not have to rely only on their neighbors.
This global information is assumed to be reliable and is treated in the same way as the neighbor information: if the signal is present and the agent experiences a small gain, it starts cooperating. 
As we can see in Fig. \ref{fig:utilI}, the signal enables the system to reach the optimum point seems independent from its size. 
This contrasts with the original system, where on average is not possible for any size to reach such high states of power utilization, regardless of the communication network topology.

\new{The second demand-side management policy is the introduction of a two-stage real time price for all agents.}
For this purpose, we have to adapt our model further.
Instead of just maximizing its own power demand, the agents now have to maximize their utility value taking the price into account. 

The price function needs to reflect the state of the system as a signal of overuse (when necessary). 
In order to build such a function, we first need to build an utility function that reflects how the power demand is valued.
We adopted the following utility from \cite{fahrioglu1999designing,Samadi2012}:
%
\[
U_{\alpha}(P_i[t],\omega_i) = 
\begin{cases} 
\omega P_i - \frac{\alpha}{2}P_i^2 & \text{if } 0 \leq P_i < \frac{\omega}{\alpha} \\
\frac{\omega^2}{2\alpha}       & \text{if } P_i \geq \frac{\omega}{\alpha}
\end{cases},
\]
where $P_i$ being the power consumption while $\alpha$ and $\omega$ determine how consumption is valued.

The function is chosen so that there exists a linear marginal benefit up to a certain point, at which the benefit does not increase anymore with higher consumption.
In our case, we keep $\alpha$ fixed at $0.2$ while $\omega_i$ is uniformly distributed between $2.05 \pm 0.05$, representing different values of consumers about the importance of power consumption. 

The price function, meanwhile, is modeled as a simple step function:
%
\[
p[t] = 
\begin{cases} 
p_1 & \text{if } n[t] \leq n^{\mathrm{opt}} \\
p_2 & \text{if } n[t] > n^{\mathrm{opt}}
\end{cases},
\]
assuming that $p_1 = 0.2$ and $p_2 = 5$.

\begin{figure}[!t]
	\centering
	\includegraphics[width=1\columnwidth]{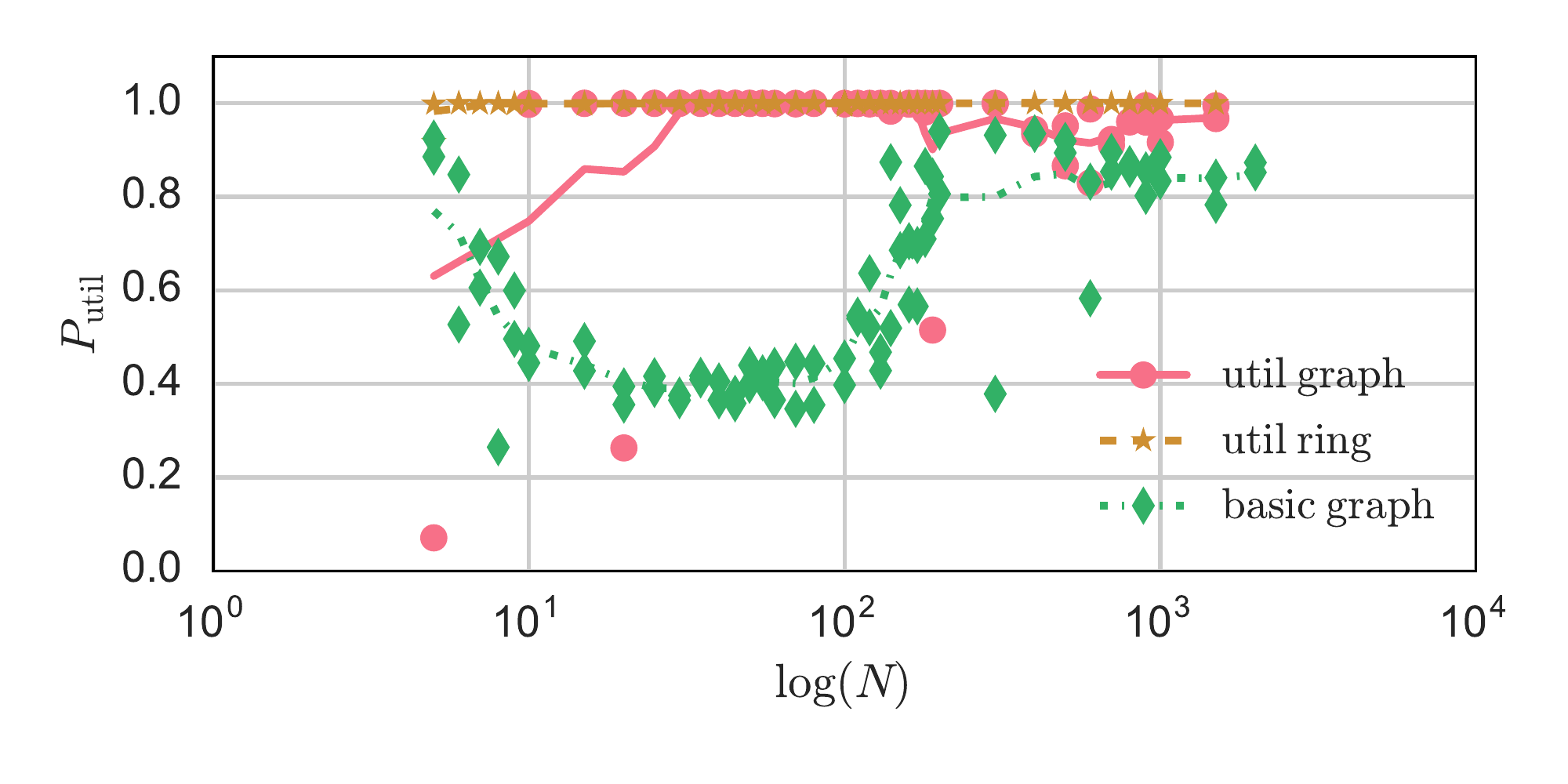}
	\caption{\label{fig:utilII}Power utilization for step pricing and agents with utility function described in this section. The green line shows the utilization for the basic model.}
	%
\end{figure}

The cost for each user is then calculated by multiplying the consumption with the current price:
\[
c_i[t] = p[t] P_i[t].
\]

The agent decision process itself is similar, but now, instead of optimizing the received power, the agents are now optimizing their benefit (the utility minus the cost):
\[
b_i[t] = U_{\alpha}(P_i[t],\omega_i) - c_i[t].
\]

The results using the proposed scheme can be viewed in Fig. \ref{fig:utilII}.
We can once again see  that the results are much better than the one presented in Section \ref{subsec:baseline}.
\new{However, if we compare the case of a simple ring graph (brown curve, labeled ``util ring'') describing the neighborhood versus a more meshed topology from Watts-Strogatz graph (green, labeled ``util graph''), the communication network (where local information about the neighbors are transmitted) has still a an influence on the final outcome.}
This effect seems to be more pronounced for small networks. 
However,  mid-sized systems, which are the most critical size when no signaling is considered, still experience stable outcomes, comparable to the global signal policy.

\section{Discussion and final remarks}
\label{sec:final}

In our previous contribution \cite{Nardelli2015a}, we showed that a simple direct current micro-grid system composed by three constituent layers (physical, informational and regulatory) may present emergent dynamics at macro-level.
This kind of very unpredictable behavior that arises is normally undesirable to manage the system and is related to a lack of global information. 
This paper shows that it is possible to counteract this behavior when some kind of global signaling based on the micro-grid aggregator. 
One  strategy is to send direct signals to end-users when the micro-grid is overused, while the other is a indirect signal through a price function.

The first strategy is a rather simple way of providing each agent with global information about micro-grid aggregate power consumption. 
This strategy enables a more stable outcome, which is much less influenced by other parameters of the simulation, like the communication network topology.
The second strategy in its turn maps the original power optimization into a cost-benefit optimization, where each agents wants to achieve the individual optimal power usage considering price and utility value of the consumed power.
While similar outputs could be demonstrated with this approach, it is necessary to point out that the choice of parameters to achieve a stable outcome in the scenario is much more complicated.
In the non-trivial cases, where the sum of the demanded power of all agents exceeds the maximum available power, the outcome is highly dependent on the choice of not only the price levels but also the valuation of the power given by the variables $\alpha$ and $\omega_i$.

These facts suggest that real time pricing in combination with demand response, which is often proposed as the way to utilize changes in available power, needs to be finely tuned in order to achieve a stable outcome. 
In other words, pricing does not inherently guarantee optimal usage and stability for the micro-grid.
Rather, the outcome of the first strategy indicates that a direct response to capacity constraints without the intermediate pricing layer is much easier to implement in a stable manner.
In any case, our results reinforce the importance of the micro-grid aggregator as a signaler of the global system state information for a stable peer relation between prosumers in the micro-grid.
We expect to further develop this simulation by adding generation capabilities in the prosumers together with a more complete market description, as indicated by our initial results \cite{kuhnlenz2016agent}.

\section*{Acknowledgments}
We would like to acknowledge the computing facilities of CSC - IT Center for Science Ltd. (Finland) that was used to run the simulation scenarios.

\bibliographystyle{IEEEtran}

\end{document}